\documentclass[pre,aps,reprint,amsmath]{revtex4-1}

\usepackage[T1]{fontenc} 
\usepackage{graphicx}
\usepackage[utf8]{inputenc}
\usepackage{lmodern}
\usepackage{bm}
\usepackage{bbold}
\usepackage[hidelinks]{hyperref}				
\usepackage[dvipsnames]{xcolor}
\usepackage{ulem}
\newcommand{\rv}{{\bf r}}

\newcommand{\fv}{{\bf f}}
\newcommand{\Jv}{{\bf J}}
\newcommand{\vel}{{\bf v}}

\newcommand{\ov}{{\boldsymbol\omega}}
\newcommand{\rot}{(\rv,\ov,t)}
\newcommand{\ro}{(\rv,\ov)}
\newcommand{\kt}{k_BT}
\newcommand{\id}{{\rm id}}
\newcommand{\eqr}[1]{Eq.~\eqref{#1}}
\newcommand{\rmint}{{\rm int}}
\newcommand{\ad}{{\rm ad}}
\newcommand{\struc}{{\rm struc}}
\newcommand{\flow}{{\rm flow}}

\begin{document}

\title{Active crystallization from power functional theory}

\author{Sophie Hermann}
\email{Sophie.Hermann@uni-bayreuth.de}
\affiliation{Theoretische Physik II, Physikalisches Institut, 
  Universit{\"a}t Bayreuth, D-95447 Bayreuth, Germany}

\author{Matthias Schmidt}
\email{Matthias.Schmidt@uni-bayreuth.de}
\affiliation{Theoretische Physik II, Physikalisches Institut, 
  Universit{\"a}t Bayreuth, D-95447 Bayreuth, Germany}
\email{Matthias.Schmidt@uni-bayreuth.de}

%%\date{20 August 2023, revised version: 25 November 2023}
\date{20 August 2023, revised version: 5 February 2024}

\begin{abstract}
We address the gas, liquid, and crystal phase behaviour of active
Brownian particles in three dimensions. The nonequilibrium force
balance at coexistence leads to equality of state functions for which
we use power functional approximations. Motility-induced phase
separation starts at a critical point and quickly becomes metastable
against active freezing for P\'eclet numbers above a nonequilibrium
triple point. The mean swim speed acts as a state variable, similar to
the density of depletion agents in colloidal demixing. We obtain
agreement with recent simulation results and correctly predict the
strength of particle number fluctuations in active fluids.
\end{abstract}

\maketitle

The occurrence of freezing in a many-body system is often due to the
presence of strong, short-ranged repulsion between the constituent
particles~\cite{hansen2013,tarazona2000prl}. Conditions of high enough
density are required for crystallization as a global ordering
phenomenon to occur and these can be induced by external constraints,
such as confinement by walls, or via interparticle attraction
\cite{leunissen2005,hueckel2020}. In colloidal systems, attraction
between the particles can be generated by adding depletion agents,
such as polymers, colloidal rods, or smaller-sized colloidal spheres.
The depletants create an effective attraction between the primary
particles and the resulting effective interaction potential is
accessible via formally integrating out (averaging over) the depletant
degrees of freedom \cite{likos2001, tuinier2003, binder2014,
  taylor2012, miyazaki2022, dijkstra1999, schmidt2000cip} and via
recent machine learning \cite{campos2021ml, campos2022ml}. In general
the resulting interaction potential has a strong many-body character,
although notable exceptions exist, such as the Asakura-Oosawa model
\cite{binder2014, miyazaki2022, dijkstra1999, schmidt2000cip}, where
for sufficiently small polymer-to-colloid size ratio a description
based on an effective pair potential is exact~\cite{dijkstra1999}.

In a striking analogy, Turci and Wilding \cite{turci2021} have
recently related the phase behaviour of three-dimensional active
Brownian particles (ABPs) \cite{turci2021, omar2021} to such
depletion-driven binary mixtures.  ABPs form a central model system
for active matter and their phase behaviour has received much prior
attention \cite{bialke2012, wysocki2014, paliwal2018,
  solon2018generalized, moore2021, speck2021pre,speck2022critical},
including the two-dimensional version of the model
\cite{bialke2012,speck2021pre,speck2022critical}. The particles
undergo overdamped Brownian motion and they self-propel (swim) along a
built-in direction, which diffuses freely. The system displays
motility-induced phase separation (MIPS) into dense and dilute
coexisting nonequilibrium steady states, despite of the absence of
explicit interparticle attraction. The phenomenon was addressed on the
basis of a wide variety of theoretical techniques \cite{paliwal2018,
  solon2018generalized, speck2021pre, speck2022critical,
  bickmann2020}, including recent work by Omar {\it et
  al.}~\cite{omar2022mips} based on forces. However, none of these
approaches has yet been applied to active freezing.

Despite the significant number of theoretical efforts
\cite{paliwal2018, solon2018generalized, speck2021pre,
  speck2022critical, bickmann2020, omar2022mips, szamel2016}, no
consensus has been reached on a common framework which would act as an
uncontested platform for the description of active systems, such as
the theory of simple liquids for spatially inhomogeneous and
phase-separated systems does in equilibrium \cite{hansen2013,
  evans1979, evans2016specialIssue, roth2010review}.  It is a rather
common point of view that ``the link between experiment and theory in
active matter is often rather qualitative''~\cite{mauleonamieva2020}.
Having a predictive theory is highly valuable though, given that much
relevant experimental work is being carried out, e.g.\ based on
light-controlled systems \cite{palacci2013}, as also used in studies
of active polarization \cite{soeker2021prl,auschra2021}, cluster
formation \cite{kuemmel2015}, the self-propulsion mechanism of Quincke
rollers \cite{mauleonamieva2020}, the experimental study of active
sedimentation \cite{ginot2018}, capillary rise
\cite{carreira2023verticalWall}, and polycrystallinity
\cite{klongvessa2019}. Equally so, simulation studies of wetting
\cite{turci2021wetting}, vortex crystal formation
\cite{james2021vortexCrystal}, inertial effects in nematic turbulence
\cite{koch2022activeTurbulence}, interfacial properties
\cite{chacon2022} and of dynamical features \cite{martinezroca2021} of
active particles could benefit from having a predictive theory.

In this Letter we use power functional theory \cite{schmidt2022rmp},
which is a general framework for the description of the dynamics of
many-body systems, including ABPs \cite{schmidt2022rmp,
  krinninger2016, krinninger2019, hermann2019pre, hermann2021molPhys,
  hermann2019prl}. We base our treatment of freezing on the active
force balance, as used in studies of active drag forces
\cite{krinninger2016,krinninger2019}, motility-induced phase
separation \cite{hermann2019pre,hermann2021molPhys} and the
interfacial tension between phase-separated states
\cite{hermann2019prl} in two-dimensional ABPs.  The theory satisfies
exact sum rules which result from Noether's theorem for correlation
functions \cite{hermann2021noether, sammueller2023whatIsLiquid} as
well as from the continuity equation for the global polarization
\cite{hermann2020polarization}.  We demonstrate that the framework
gives a physically sound and quantitatively reasonable account of the
full phase behaviour of ABPs in three dimensions. In their analogy,
Turci and Wilding \cite{turci2021} suggest that the P\'eclet number,
which measures the strength of the self-propulsion in the active
system relative to diffusive motion, is akin to the depletants'
fugacity (or polymer reservoir density) in an equilibrium mixture.  We
confirm and extend this point of view, as in our theoretical approach
the mean swim speed plays a role akin to the actual polymer density in
the colloid-polymer mixture.

We work on the level of one-body correlation functions, which depend
on position $\rv$ and on particle orientation, as represented by a
unit vector $\ov$. The continuity equation relates the divergence of
the translational current $\Jv\rot$ and of the rotational current
$\Jv^\omega\rot$ to temporal changes of the one-body density
distribution according to:
\begin{align}
  \frac{\partial \rho\rot}{\partial t} 
  &= - \nabla \cdot \Jv\rot
  - \nabla^\omega \cdot \Jv^\omega\rot.
  \label{EQcontinuity}
\end{align}
Here $\nabla$ and $\nabla^\omega$ indicate the derivatives with
respect to~$\rv$ and $\ov$, respectively, and the density profile
$\rho\rot$ is position- and orientation-resolved.  We consider steady
states such that the left hand side of \eqr{EQcontinuity} vanishes and
we drop the time argument $t$ from here on.  As no explicit torques
act in the system, the orientational current stems solely from the
free rotational diffusion of the active spheres:
$\Jv^\omega\ro=-D_{\rm rot} \nabla^\omega \rho\ro$, where $D_{\rm
  rot}$ indicates the rotational diffusion constant. For the present
case of overdamped active motion, the exact force balance is given by:
\begin{align}
  \gamma \vel(\rv,\ov) 
  &= \fv_{\rm id}(\rv,\ov)
  + \fv_{\rm int}(\rv,\ov)
  +\gamma s \ov.
  \label{EQforceBalance}
\end{align}
The left hand side of \eqr{EQforceBalance} represents the negative
friction force with friction constant $\gamma$ and the velocity field
is the ratio of current and density, $\vel\ro=\Jv\ro/\rho\ro$. The
three driving contributions on the right hand side of
\eqr{EQforceBalance} are the ideal diffusive force field $\fv_\id\ro =
-\kt \nabla\ln\rho\ro$, the internal force field $\fv_{\rm int}\ro$,
which arises from the Weeks-Chandler-Anderson (WCA) interparticle
interactions, and the swim force $\gamma s \ov$ with~$s$ indicating
the speed of free swimming.  The one-body interparticle interaction
force field $\fv_\rmint\ro$ is accessible via sampling in simulations
\cite{krinninger2016, krinninger2019, hermann2019pre,
  hermann2021molPhys} and via machine-learning, as recently
demonstrated in passive flow \cite{delasheras2023perspective} and in
equilibrium \cite{sammueller2023neural,
  sammueller2023neuralTutorial}. When averaged over orientation~$\ov$
there is no net flow in the stationary states considered here, $\int
d\ov\Jv(\rv,\ov)=0$.

We split the interparticle forces according to~\cite{hermann2019prl,
  delasheras2020fourForces}:
\begin{align}
  \fv_\rmint\ro &= \fv_\ad(\rv) + \fv_\flow\ro + \fv_\struc\ro,
  \label{EQsupSplitting}
\end{align}
where the right hand side consists of the adiabatic force field
$\fv_\ad(\rv)$, the superadiabatic flow force field $\fv_\flow\ro$ and
the superadiabatic structural force field $\fv_\struc\ro$.  Here the
adiabatic force field $\fv_\ad(\rv)$ is defined as acting in an
equilibrium system of passive WCA particles that do not swim.  The WCA
particles are spheres and hence there is no nontrivial dependence on
$\ov$ in the adiabatic system. Its density distribution
$\bar\rho(\rv)$ is identical to the orientation-integrated density
distribution in the active system. In the adiabatic system
$\bar\rho(\rv)$ is stabilized by an external potential.

If one wishes to think in terms of functional dependencies, then
$\fv_\ad(\rv)$ is an instantaneous density functional, in the sense of
functional dependencies as they form the core of classical density
functional theory of inhomogeneous liquids and solids \cite{evans1979,
  hansen2013, evans2016specialIssue, roth2010review, schmidt2022rmp}.
Both the flow and the structural force contributions in
\eqr{EQforceBalance} are of superadiabatic nature, i.e.\ they are
genuine nonequilibrium force fields which arise from the interparticle
interactions \cite{schmidt2022rmp}.  The flow and structural
nonequilibrium forces, $\fv_\flow\ro$ and $\fv_\struc\ro$, have
characterizing symmetry properties under motion reversal. Here
$\fv_\flow\ro$ changes its sign under sign change of the steady state
velocity profile while $\fv_\struc\ro$ remains unaffected by the same
transformation \cite{delasheras2020fourForces, schmidt2022rmp,
  delasheras2023perspective}.  In equilibrium, as well as in passive
uniaxial flow, the three force contributions were shown to be amenable
to supervised machine learning \cite{delasheras2023perspective,
  sammueller2023neural, sammueller2023neuralTutorial}, which we take
as confirmation of the general force splitting concept
\eqref{EQsupSplitting}, as is here applied to the active system.

The flow force $\fv_\flow\ro$, as it is part of
Eq.~\eqref{EQsupSplitting}, is defined to compensate the friction and
the active force in the force balance relationship
\eqref{EQforceBalance} such that equality is achieved:
\begin{align}
  \gamma \vel(\rv,\ov) &=
  \fv_{\rm flow}(\rv,\ov) + \gamma s \ov.
  \label{EQflowBalance}
\end{align}
The flow equation \eqref{EQflowBalance} is invariant under motion
reversal \cite{schmidt2022rmp, delasheras2020fourForces,
  delasheras2023perspective} and it affects the spatial structure
formation as represented by the density profile, only indirectly, as
we detail below.  As an approximation we resort to the superadiabatic
drag force of Ref.~\cite{krinninger2016}, which in bulk has the simple
form $\fv_{\rm flow}(\ov) =-\gamma v_b \ov \rho_b/(\rho_j-\rho_b)$,
where $v_b = \vel\cdot\ov$ is the mean forward swim speed
\cite{solon2018generalized, speck2021pre, speck2022critical,
  bickmann2020}, which is reduced due to particle collisions, when
compared to the free swim speed $s$. The assumption yields the common
linear relationship of the mean swim velocity and the average density,
$v_b/s=1-\rho_b/\rho_j$, independent of position and orientation. The
parameter $\rho_j=\rm const$, which we adjust empirically, determines
the slope of the decay of $v_b$ with bulk density and we adjust its
value to $\rho_j=1.436\sigma^{-3}$, where $\sigma$ is the lengthscale
of the WCA pair potential, in order to approximate the observed
behaviour \cite{vandamme2019}.

The superadiabatic structural force field $\fv_{\rm sup}\ro$ balances
the remaining adiabatic and ideal terms in \eqr{EQforceBalance}, which
implies the following force cancellation:
\begin{align}
  0 &= \fv_{\rm id}(\rv) + \fv_{\rm ad}(\rv) + \fv_{\rm struc}(\rv).
  \label{EQstrucBalance}
\end{align}
As a consistency check, the sum of Eqs.~\eqref{EQflowBalance} and
\eqref{EQstrucBalance} recovers the full force balance
relationship~\eqref{EQforceBalance}. The ideal term is generally
numerically small, and we hence approximate the exact ideal force
$-\kt\nabla\ln\rho\ro \approx -\kt\nabla\ln\bar\rho(\rv)\equiv
\fv_\id(\rv)$, where as before $\bar\rho(\rv)$ is the
position-dependent and orientation-averaged one-body density profile.
Equation \eqref{EQstrucBalance} balances the repulsion that acts in
the adiabatic system with the nonequilibrium force contributions. We
recall that the adiabatic system consists of steeply repulsive spheres
without orientations.  Hence the structural nonequilibrium forces
necessarily need also be independent of orientation, $\fv_{\rm
  struc}(\rv)$, in order to satisfy \eqr{EQstrucBalance}.

As all force fields in \eqr{EQstrucBalance} are of gradient nature
[the non-gradient forces are contained in \eqr{EQflowBalance}], we can
integrate in position and obtain the following chemical potential
balance:
\begin{align}
  \mu_\id(\rv) + \mu_\ad(\rv) + \mu_{\rm struc}(\rv) &= \mu.
  \label{EQmuBalance}
\end{align}
Here $\mu=\rm const$ arises from the spatial integration. All terms on
the left hand side of \eqr{EQmuBalance} are solely defined by
generating via spatial differentiation the (negative) force
contributions that occur in the structural force
balance~\eqref{EQstrucBalance}. Explicitly, we have
$\fv_\id(\rv)=-\nabla \mu_\id(\rv)$, with the standard ideal gas
chemical potential expression: $\mu_\id(\rv)= -\kt\ln\bar\rho(\rv)$;
the adiabatic force field: $\fv_\ad(\rv)=-\nabla\mu_\ad(\rv)$; and the
superadiabatic structural force field:
$\fv_\struc(\rv)=-\nabla\mu_\struc(\rv)$.

Up to having neglected the orientation-dependence of the ideal gas
contribution and the assumption of the specific simple form of
$\fv_\flow(\ov)$, the framework thus far developed is exact and we
have to resort to approximations to make further progress. We first
turn to the adiabatic contribution.  The adiabatic state is simply the
equilibrium WCA model, which per se has no gas-liquid coexistence due
to its lack of interparticle attraction. Treating fluid states of
repulsive spheres is straightforward. We approximate the system by
hard spheres and use a modified Carnahan-Starling equation of state
\cite{paricaud2015saft}, which correctly accounts for the behaviour at
very high densities, as is relevant for ABPs in the parameter regime
considered here. The corresponding bulk excess free energy $A_\ad$ is
given by
\begin{align}
\frac{A_\ad}{Nk_BT} &=  \frac{3\eta}{1-\eta} \notag\\
&+ \eta \left\{(1-\eta)\left[(1-\eta\left(1+\frac{1-\eta_j}{\eta_j}
  {\rm e}^{b(\eta-\eta_j)}\right)\right] \right\}^{-1}
\end{align}
where $\eta = \pi \sigma^3 \rho_b /6$ is the packing fraction,
$\eta_j=0.655$ is the densest possible packing fraction in this
approximation and setting $b=50$ is an empirical choice
\cite{paricaud2015saft}.

Due to the more complex three-dimensional system we here choose an
advanced approximation for the adiabatic fluid equation of state, as
compared to previous work in two dimensions \cite{hermann2019prl,
  hermann2019pre, hermann2021molPhys}, where the simple
scaled-particle theory was sufficient. An analytical expression for
the bulk chemical potential in the adiabatic system then follows from
the standard identity $\mu_\ad^b(\rho_b) = [A_\ad + \eta \partial
  A_\ad/\partial \eta]/N$. We use a local density approximation where
we evaluate the bulk expression at the value of the local density
profile, i.e.\ $\mu_\ad(\rv)=\mu_\ad^b(\bar\rho(\rv))$.

In order to approximate the equation of state of the adiabatic crystal
we resort to the cell theory \cite{hoover1967, hoover1968,
  frenkel1984, royall2023review}. This yields the chemical potential of the fcc crystal
as $\beta\mu_{\rm cell}= \ln(\sqrt{2}) + 3 \ln(\Lambda/\sigma) - 3
\ln(\xi-1) + \xi/(\xi-1)$, where $\xi=[\pi \sqrt{2}/(6 \eta)]^{1/3}$
and $\Lambda$ is the thermal de Broglie wavelength which we set to
$\Lambda=\sigma$.  We only take account of the mean crystal density,
and set $\mu_\ad + \mu_\id =\mu_{\rm cell}$ for the treatment of the
crystalline phase.

The remaining task is to approximate the superadiabatic structural
chemical potential contribution, $\mu_\struc(\rv)$. Here we resort to
the ``quiet life'' approximation, which was successfully used to
describe active gas-liquid phase separation in two dimensions, along
with the force balance across the free interface between the active
bulk states \cite{hermann2019pre,hermann2021molPhys}. This
approximation takes into account, in arguably the simplest correct
way, the dependence on both the local density and on the local
velocity.  As the force is structural, it is necessarily even in
powers of the velocity. A simple choice which is linear in density and
quadratic in velocity \cite{hermann2019pre, hermann2021molPhys} reads
as:
\begin{align}
  \mu_\text{struc}(\rv) &= 
  \frac{e_1\gamma}{6 D_\text{rot}} 
  v^2(\rv) \frac{\bar{\rho}(\rv)}{\rho_j},
  \label{EQmuStruc}
\end{align}
where $e_1= 0.285$ is a constant that determines the overall strength.
This parameter plays the role of a fundamental transport coefficient
of the ABP system. Here we fix its value empirically and leave a
first-principles derivation to future work.  As we also only address
bulk states we set the (squared) local swim velocity to its mean bulk
value, $v^2(\rv)=v_b^2$.  Crucially, we use the same
approximation~\eqref{EQmuStruc} for $\mu_\struc$ in all three phases
and in particular the same value of the constant $e_1.$

\begin{figure}
  \includegraphics[width=0.99\columnwidth,angle=0]{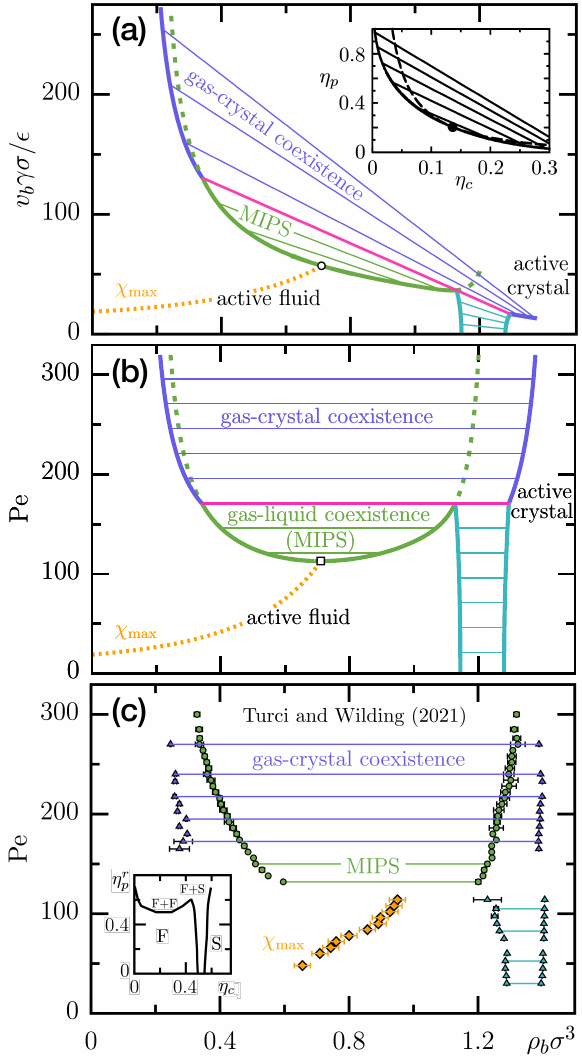}  
\caption{Phase diagram for three-dimensional ABPs. (a) Theoretical
  result as a function of the scaled bulk density $\rho_b \sigma^3$
  and the scaled average swim speed $v_b\gamma \sigma/\epsilon$.
  Shown are stable (solid lines) and metastable (dashed lines)
  binodals; slanted tielines connect coexisting nonequilibrium states.
  The orange dotted line indicates the line of maximal compressibility
  $\chi_{\rm max}$.  Note the similarity to a phase diagram of a
  colloid-polymer mixture (inset, adapted from
  Ref.~\cite{schmidt2000cip}) as a function of the colloid (polymer)
  packing fraction $\eta^C (\eta^P)$.  (b) Same as (a) but shown as a
  function of $\rho_b\sigma^3$ and the P\'eclet number Pe. The tie
  lines are horizontal in this representation. (c) Same as (b) but
  obtained from computer simulations in Ref.~\cite{turci2021}. Shown
  are active gas-active fluid (circles), active gas-crystal
  (triangles) coexistence densities as well as $\chi_{\rm max}$
  (diamonds). The inset is a schematic phase diagram for a
  polymer-colloid mixture with size ratio $q=0.6$, taken from
  Ref.~\cite{dijkstra1999}.}
  \label{FIGpd}
\end{figure}

Nonequilibrium phase coexistence is obtained via the mechanical
balance of the total force, which in our framework implies equality of
the values of the chemical potential, see \eqr{EQmuBalance}, in the
coexisting phases, as well as the equality of the pressure.  The
pressure is obtained from integrating the standard relation $ \rho_b
\partial \mu(\rho_b)/\partial \rho_b = \partial P(\rho_b)/\partial
\rho_b$. The resulting phase diagram is shown in Fig.~\ref{FIGpd} as a
function of the bulk density $\rho_b$ and either the mean swim speed
$v_b$ (a) or the free swim speed $s$ (b), as expressed in scaled form
by the P\'eclet number ${\rm Pe}=s \sigma\gamma/(k_BT)$. The topology
of the phase diagram matches that obtained in simulation work
\cite{turci2021,omar2021} and the reasonable agreement with the
simulation results by Turci and Wilding \cite{turci2021} is very
satisfactory; their results for the phase diagram are displayed in
Fig~\ref{FIGpd}(c).  Our theory reproduces the marginal stability
\cite{omar2021} of active gas-liquid coexistence with respect to
freezing into a dense fcc crystal. The coexisting gas has relatively
high density, in stark contrast to the strong dilution of the
coexisting gas that occurs quickly in equilibrium phase separation
when moving away from the triple point.

\begin{figure}
  \includegraphics[width=0.95\columnwidth,angle=0]{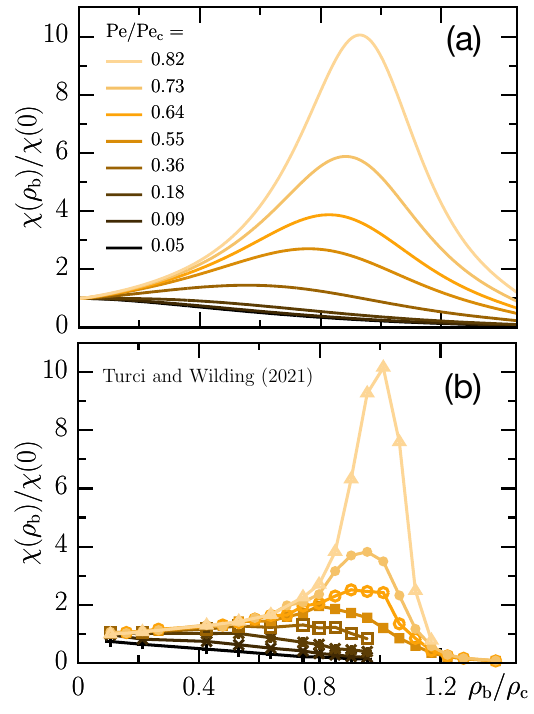}
  \caption{Scaled compressibility $\chi(\rho_b)/\chi(0)$ as a function
    of the scaled bulk density $\rho_b/\rho_c$, where $\rho_c$ is the
    density at the MIPS critical point and $\chi(0)$ is the
    low-density limit of the compressibility. The theoretical results
    in panel (a) are obtained from differentiating
    $\chi(\rho_b)=\partial\rho(\mu)/ \partial \mu|_{T,\rm Pe}$, where
    $\mu$ is the (total) nonequilibrium chemical potential. Results
    are shown for a sequence of P\'eclet numbers (as indicated),
    scaled by the value at the critical point. Panel (b) shows the
    corresponding simulation results of Ref.~\cite{turci2021}; for the
    purpose of this comparison, we take ${\rm Pe}_c = 36$, $\rho_c =
    0.94$ in simulation \cite{turci2021} and ${\rm Pe}_c = 37.6$,
    $\rho_c = 0.71$ for the theory. The lines in (b) connect the data
    points to guide the eye.}
  \label{FIGchi}
\end{figure}

On the basis of the similarity of their simulation results for the
active system to depletion-induced phase behaviour in equilibrium
(comparing the main plot and inset of Fig.~\ref{FIGpd}(c)), Turci and
Wilding \cite{turci2021} draw conclusions about the presence and
relevance of effective many-body interactions that govern the active
system.  While it is well-established that in active systems the
P\'eclet number plays a role similar to that of temperature in
equilibrium, the proposal by Turci and Wilding leaves open whether one
should think of the activity as only generating many-body effects that
are akin to those of depletants or whether the active system contains
actual degrees of freedom that have not been properly appreciated.

Based on the success of our theory, we argue that the latter is the
case and that besides the density, the velocity field is an intrinsic
degree of freedom that the nonequilibrium system can regulate freely
and self-consistently. To demonstrate the validity of this concept, we
use the actual mean velocity $v_b$ instead of Pe as a state variable
in Fig.~\ref{FIGpd}(a). The swim speed $v_b$ is high in the coexisting
gas, low in the coexisting liquid, and even lower in the coexisting
crystal. The latter property is consistent with Caprini {\it et
  al.}\ reporting very low swim speeds in solid clusters of the
two-dimensional ABP system, see the Supplemental Material of
Ref.~\cite {caprini2020}. This behaviour is analogous to what is found
in depletion-driven phase separation, when going from the reservoir
density of the depletant to the actual depletant density in the
system~\cite{dijkstra1999, schmidt2000cip}.  The observed similarity
in the form of the phase diagram is striking, compare the main plot
and the inset of Fig.~\ref{FIGpd}(a).

We next investigate whether our proposed theory is predictive beyond
the phase diagram. In their simulation work \cite{turci2021}, Turci
and Wilding have investigated the statistics of particle number
fluctuations, as they occur in small virtual subboxes of the global
system.  The strength of fluctuations $\chi(\rho_b)$ is taken to be a
proxy for the compressibility, as can in equilibrium be obtained from
the thermodynamical derivative $\partial\rho_b(\mu)/\partial\mu$,
carried out in the grand ensemble where global particle number
fluctuations occur. These fluctuations are absent in the present
system, as the particle number is conserved in time [we recall the
  validity of even the locally resolved continuity equation
  \eqref{EQcontinuity}].

Within our nonequilibrium framework the partial derivative
$\chi(\mu_b)=\partial \rho_b(\mu)/\partial \mu$ is well-defined. Here
$\mu$ is the total chemical potential and we recall its
splitting~\eqref{EQmuBalance} into adiabatic and superadiabatic
contributions. We invert via $\chi(\mu_b) = [\partial
  \mu(\rho_b)/\partial\rho_b]^{-1}$ with the derivative taken at
$T,{\rm Pe}={\rm const}$. We normalize with respect to the low density
behaviour, $\chi(\rho_b)/\chi(0)$, as has also been done in the
simulations \cite{turci2021}. In order to create further common ground
we scale the density axis by the respective value of the critical
density $\rho_c$.  From the setup of the theory, we expect
$\chi(\rho_b)/\chi(0)$ to be a measure of particle fluctuations and we
show numerical results in Fig.~\ref{FIGchi}(a) as a function of
$\rho_b/\rho_c$ for a range of different values of ${\rm Pe}/{\rm
  Pe}_c < 1$, where ${\rm Pe}_c$ indicates the critical value of the
Peclet number. We find that the theory produces the same bell-shaped
variation upon increasing density at fixed Pe, as is apparent in the
simulation results reproduced in Fig.~\ref{FIGchi}(b). The maximum
becomes much more pronounced upon increasing ${\rm Pe}/{\rm Pe}_c$ and
the theoretical prediction consistently diverges at the nonequilibrium
critical point. The position of the maximum of $\chi(\rho_b)/\chi(0)$
traces a line in the phase diagram. The result is shown in
Fig.~\ref{FIGpd}, which again agrees very well with the simulation
data (compare the orange line in Fig.~\ref{FIGpd}(b) with the orange
symbols in panel in Fig.~\ref{FIGpd}(c)).

We take this satisfactory agreement of the simulation
results~\cite{turci2021} for particle number fluctuations against an
analogous parametric derivative of our nonequilibrium state function
as a test of the intrinsic consistency of our treatment. The theory in
particular reproduces the strong increase in measured particle number
fluctuations near the nonequilibrium critical point. As we derive
nonequilibrium phase coexistence from the same state function, we
conclude that our approach indeed captures much of the essence of the
nonequilibrium statistical physics under consideration.
We recall that we obtain the bulk pressure $P(\rho_b)$ via integrating
the identity $\rho_b \partial \mu(\rho_b)/\partial \rho_b = \partial
P(\rho_b)/\partial \rho_b$, which implies mechanical stability in our
formulation of the nonequilibrium physics~\cite{hermann2019pre,
  hermann2021molPhys, hermann2019prl}.  The investigation of the
particle number fluctuations, as measured via $\chi(\rho_b)$, provides
a test for the validity of the ``source'' material on the left hand
side of the equation, as $\rho_b \partial \mu(\rho_b)/\partial \rho_b
=\rho_b/\chi(\mu_b)$.
To shed further light on the particle number fluctuation problem,
given the prominent role that the Ornstein-Zernike theory plays in
describing fluctuations inequilibrium \cite{hansen2013}, we can
imagine that concepts of the nonequilibrium Ornstein-Zernike
framework~\cite{brader2013noz, brader2014noz} could pave a way
forward.

In summary we have investigated the nonequilibrium phase
behaviour of ABPs in three dimensions, based on power functional
concepts.  The central assumption is that the formally exact
nonequilibrium force balance relationship contains a nonequilibrium
structural force contribution, as obtained by the negative spatial
gradient of a corresponding superadiabatic chemical
potential,~\eqr{EQmuStruc}.  We have shown that the theory predicts
the phase diagram correctly and that nonequilibrium particle number
fluctuations are described in agreement with the observations in
simulations.  We envisage that going beyond the simple cell theory for
the description of the crystal is possible with classical density
functional theory \cite{haering2015} based on fundamental measure
theory \cite{tarazona2000prl, roth2010review} as used to study the
direct correlation function in crystals \cite{lin2021dcf}.  Given the
recent progress in measurement of intercolloidal forces in gel states
\cite{dong2022}, it seems not inconceivable that experiments can shed
further light on active forces.

In our present treatment we have characterized the steady-state
one-body velocity field $\vel(\rv,\ov)$ in each of the three
nonequilibrium phases by the value of the mean swim speed $v_b$. For
both the active gas and the active liquid phase, due to their
rotational and translational invariances, $v_b=\vel(\rv,\ov)\cdot\ov$
indeed becomes independent of position $\rv$ and of orientation~$\ov$.
Thus the full information is retained and the velocity field is
$\vel(\rv,\ov)=v_b \ov$ for active bulk fluids. We also describe the
crystal on the basis of the single parameter~$v_b$, which we take to
be a coarse-grained and global measure of the activity across the
spatial inhomogeneity of the lattice.  It remains thus to study and to
describe fully the inhomogeneous flow field in the crystal,
$\vel(\rv,\ov)$, which we deem to be an interesting problem, not least
in the light of the velocity-alignment effects identified by Caprini
{\it et al.}  in two dimensions \cite{caprini2020}.

In very recent work Evans and Omar \cite{evans2023activeFreezing} have
addressed active freezing theoretically. Furthermore an experimental
investigation of freezing of passive colloids in an active solvent was
reported by Massana-Cid {\it et al.}~\cite{massanacid2024}. It would
be interesting to relate to these studies in future work.

\acknowledgments We thank Francesco Turci for sending us the
simulation data of Ref.~\cite{turci2021} and him, Nigel Wilding, and
Daniel de las Heras for useful and inspiring discussions.

\end{document}